\documentclass{aa}  

\usepackage{graphicx}
\usepackage{txfonts}
\usepackage{natbib}
\usepackage{amsmath} 
\usepackage{amssymb}
\usepackage{mathtools}
\usepackage{microtype}
\usepackage[dvipsnames]{xcolor}
\definecolor{CiteRed}{RGB}{110, 0, 0}
\usepackage{multirow}
\usepackage{array}
\usepackage{makecell}
\usepackage{siunitx}
\usepackage{xcolor}
\usepackage{orcidlink}
\usepackage{graphicx}
\usepackage{subcaption}
\usepackage{verbatim}
\graphicspath{ {img/} {plot/} {fig/} }
\defcitealias{SternSvensson96}{SS96}
\defcitealias{Bazzanini24}{B24}

\usepackage{hyperref}
\hypersetup{
    breaklinks,
    colorlinks,
    citecolor=CiteRed,
    linkcolor=NavyBlue,
    urlcolor=NavyBlue,
    linktoc=page,
    pdftitle={},
    pdfkeywords={GRBs, Machine Learning, Genetic Algorithms},
    pdfauthor={Manuele Maistrello},
    pdfcreator={\LaTeX}
}

\begin{document} 

   \title{An advanced pulse-avalanche stochastic model of long gamma-ray burst light curves}
   \author{M.~Maistrello~\inst{1,2}\fnmsep\thanks{\texttt{mstmnl[at]unife[dot]it}}\orcidlink{0009-0000-4422-4151}\and
           L.~Ferro~\inst{1,2,4}\orcidlink{0009-0006-1140-6913}\and
           L.~Bazzanini~\inst{1,2}\orcidlink{0000-0003-0727-0137}\and
           R.~Maccary~\inst{1,2}\orcidlink{0000-0002-8799-2510}\and
           C.~Guidorzi~\inst{1,2,3}\orcidlink{0000-0001-6869-0835}
          }
          
   \institute{
   Department of Physics and Earth Science, University of Ferrara, via Saragat 1, I--44122, Ferrara, Italy\label{unife}\and 
   INAF -- Osservatorio di Astrofisica e Scienza dello Spazio di Bologna, Via Piero Gobetti 101, I-40129 Bologna, Italy\label{oabo}\and 
   INFN -- Sezione di Ferrara, via Saragat 1, I--44122, Ferrara, Italy\label{infnfe} \and
   Deutsches Elektronen-Synchrotron DESY, Platanenallee 6, 15738 Zeuthen, Germany\label{desy}
   }

  \date{Received xxx; accepted xxx}

  \abstract
   {A unified explanation of the variety of long-duration gamma-ray burst (GRB) light curves (LCs) is essential for identifying the dissipation mechanism and possibly the nature of their central engines. In the past, a model was proposed to describe GRB LCs as the outcome of a stochastic pulse avalanche process, possibly originating from a turbulent regime, and it was tested by comparing average temporal properties of simulated and real LCs. Recently, we revived this model and optimised its parameters using a genetic algorithm (GA), a machine-learning-based approach. Our findings suggested that GRB inner engines may operate near a critical regime.}
   {Here we present an advanced version of the model, which allows us to constrain the peak flux distribution of individual pulses, and evaluate its performance on a new dataset of GRBs observed by the \emph{Fermi} Gamma-ray Burst Monitor (GBM).}
   {After introducing new model parameters and a further comparison metric, that is the observed signal-to-noise (S/N) distribution, we test the new model on three complementary datasets: CGRO/BATSE, Swift/BAT, and Fermi/GBM. As in our previous work, the model parameters are optimised using a GA.}
   {The updated sets of parameters achieve a further reduction in loss compared to both the original model and our earlier optimisation. The different values of the parameters across the datasets are shown to originate from the different energy passbands, effective areas, trigger algorithms, and, ultimately, different GRB populations of the three experiments.}
   {Our results further underpin the stochastic and avalanche character of the dissipation process behind long GRB prompt emission, with an emphasis on the near-critical behaviour, and establish this new model as a reliable tool for generating realistic GRB LCs as they would be seen with future experiments.}

   \maketitle

\section{Introduction}
\label{sec:intro}

Most long-duration gamma-ray bursts (LGRBs) are believed to originate from relativistic jets launched by the newborn compact object --either an hyper-accreting black hole (BH) or a millisecond magnetar-- resulted from the core collapse of certain hydrogen-stripped massive stars, which might either be single (so-called ''collapsar'', \citealt{Woosley93, Paczynski98, MacFadyen99}) or belong to a binary system \citep{FryerHeger05, RuedaRuffini12, Chrimes20}\footnote{There is a handful of exceptional cases of LGRBs that are likely to have a binary compact object merger origin.}.

Deciphering the complexity and diversity of LGRB light curves (LCs) is crucial to understand the dissipation mechanism into gamma-rays and, possibly, to constrain the nature of the engines powering these extraordinary phenomena. One possibility is that the observed variety reflects the imprint of the inner engine activity on the relativistic jet, coupled with the effects of the jet propagation through the progenitor's stellar envelope (see \citealt{Gottlieb20a, Gottlieb20b, Gottlieb21a, Gottlieb21b} for state-of-the-art simulations). Alternatively, runaway magnetic reconnection cascades occurring at larger radii could be the source of short-timescale variability \citep{ICMART}. Despite the significant progress in understanding the energetics, structure, and composition of the jets, as well as the role of the magnetic field, a unified model that accounts for the variety of the LC properties, such as the distributions of the number of peaks per GRB, of the energy of each individual pulse, and of waiting times, is still missing.

In this context, determining whether the dynamics of the inner engine can be described as either non-linear and deterministic or stochastic is already a key issue. Almost two decades ago, \citet[hereafter SS96]{SternSvensson96} proposed a stochastic framework for modelling the inner engine activity, specifically how energy is released over time. The model, based on a pulse avalanche process working near a critical regime, suggests that primary episodes of energy release could trigger secondary episodes, continuing until the process becomes sub-critical. With an educated guess of the model parameters, \citetalias{SternSvensson96} successfully generated LCs that reproduced several average temporal properties of GRBs observed by the Burst And Transient Source Experiment (BATSE) aboard former Compton Gamma-Ray Observatory (CGRO; 1991--2000). The stochastic and pulse-avalanche character of the model could hint to a turbulent regime of the emitting fluid or to some kind of instability (e.g., magneto-rotational or gravitational) developing in the hyper-accreting disc.

In our previous work \citep[hereafter B24]{Bazzanini24}, we revived the \citetalias{SternSvensson96} model and optimised its parameters using a genetic algorithm (GA). Additionally, we further tested the model for the first time with an independent sample of GRBs detected by the Burst Alert Telescope (BAT; \citealt{Barthelmy05}) aboard the Neil Gehrels Swift Observatory \citep{Gehrels04}. Our findings reinforced the conclusions of \citetalias{SternSvensson96}, showing that a relatively simple pulse avalanche model can reproduce realistic synthetic LCs, whose average properties match those observed in real BATSE and Swift/BAT data. Furthermore, we provided compelling evidence supporting the stochastic nature of LGRB engines, which must operate close to a critical regime.   

In this paper, we propose an advanced version of the \citetalias{SternSvensson96} model, which reproduces a realistic signal-to-noise ratio (S/N) distribution for each dataset by modelling the $\log N-\log F$ distribution of individual pulses.
Following \citetalias{Bazzanini24}, we optimised the model parameters with a GA and applied it for the first time to a new sample of GRBs observed by the Fermi Gamma-ray Burst Monitor (GBM; \citealt{Meegan09}). This work is organised as follows. We describe the sample selection and the metrics used for the model optimisation in Section~\ref{sec:data}. The model, along with the implementation of the GA, is presented in Section~\ref{sec:methods}. The results and conclusion are reported in Sections~\ref{sec:results} and~\ref{sec:disc_conc}, respectively.

\section{Data analysis}
\label{sec:data}

\subsection{Sample selection}
\label{ss:lc_selection}

Following \citetalias{Bazzanini24}, our first dataset comprises GRBs from the 4B BATSE catalogue \citep{Paciesas99}, referred to as the BATSE sample. The second dataset includes GRBs detected by Swift/BAT between January 2005 and November 2023, named as the Swift sample. For both datasets, we used background-subtracted LCs in their respective total passbands (25--2000~keV for BATSE and 15--150~keV for Swift), with a time resolution of 64~ms.

We focused exclusively on LGRBs, requiring $T_{90} > 2$~s and excluded the cases for which compelling evidence for a compact merger origin was established (GRB\,060614, GRB\,211211A, GRB\,191019A). As in \citetalias{Bazzanini24}, we employed the $T_{20\%}$ interval, defined as the time during which the LC signal exceeds 20\% of its maximum intensity, as a proxy for duration. LCs data outside $5/3 \times T_{20\%}$ were zero-padded to exclude non-significant regions. While \citetalias{Bazzanini24} required post-peak durations of at least 150~s, the updated procedure allows for relaxed constraints while maintaining the temporal integrity of the curves. Additionally, the signal-to-noise (S/N) thresholds were reduced from 70 (15) for BATSE (Swift) to 15 (10), enabling the inclusion of GRBs with lower statistical quality. This adjustment increased the diversity and completeness of the datasets in terms of GRB morphologies. Applying these criteria, the initial samples' sizes of 2024 and 1389 GRBs for BATSE and Swift, respectively, shrank to 1155 and 635 GRBs respectively, which will be hereafter referred to as the BATSE and the Swift samples.

A third dataset, including 2356 GRBs from the fourth Fermi/GBM catalogue (July 2008--July 2018, \citealt{vonKienlin20}), was also included. We used background-subtracted LCs in the full NaI scintillators passband (8--1000~keV) with a bin time of 64~ms. Background subtraction was performed using publicly available GBM Tools\footnote{\url{https://fermi.gsfc.nasa.gov/ssc/data/analysis/gbm/gbm_data_tools/gdt-docs/}} following the procedure described in \citet{Maccary24}. We selected 1120 GRBs from the Fermi sample using the same criteria as for BATSE and Swift, with a threshold of $\mathrm{S/N} > 15$. Hereafter, this will be referred to as the Fermi sample.

\subsection{Statistical metrics}
\label{ss:metrics}

The degree of similarity between the simulated and real LCs was assessed using the following metrics:
\begin{enumerate}
    \item The average post-peak time profile, aligned with the brightest peak, over the first 150~s, $\langle F / F_p \rangle$, where $F_p$ is the peak count rate.
    \item The average third moment of the post-peak time profile, $\langle (F / F_p)^3 \rangle$.
    \item The average auto-correlation function (ACF).
    \item The $T_{20\%}$ distribution.
    \item The S/N distribution of the LCs.
\end{enumerate}
The first four metrics are as defined in \citetalias{Bazzanini24} (see references therein for more details). The fifth metric was introduced to ensure a realistic intensity distribution (see Sec.~\ref{ss:lc_selection}).

\section{Methods}
\label{sec:methods}

\subsection{Stochastic pulse avalanche model}
\label{ss:model}

The model used to simulate the LCs was originally proposed by \citetalias{SternSvensson96} and recently optimised by \citetalias{Bazzanini24} using a GA. Below, we summarise the main features of the model and algorithm, referring to the original works for more detailed explanations. 

The \citetalias{SternSvensson96} model is based on the following assumptions:
\begin{enumerate}
    \item Each LC is a unique random realisation of a common stochastic process, specifically a pulse avalanche, with parameters confined to narrow ranges. 
    \item The process is scale-invariant in time.
    \item The process operates near its critical regime.
\end{enumerate}
In this framework, spontaneous primary pulses trigger secondary pulses in a recursive manner until the system transitions to a subcritical regime. The superposition of primary (parent) and secondary (child) pulses makes up the GRB LC. Each pulse follows a Gaussian rise and exponential decay \citep{Norris96}:
\begin{equation}
    F(t) =
    \begin{cases}
        A \exp[-(t-t_p)^2/\tau_r^2] & \text{for $t < t_p$,} \\
        A \exp[-(t-t_p)/\tau]       & \text{for $t > t_p$,} \\
    \end{cases}
\end{equation}
where $A$ is the peak flux amplitude, $t_p$ the peak time, $\tau$ the pulse time constant (approximately equal to the pulse width), and $\tau_r = \tau/2$. Originally, \citetalias{SternSvensson96} sampled $A$ from a uniform distribution $\mathcal{U}[0, 1]$, whereas for each GRB \citetalias{Bazzanini24} sampled $A$ from $\mathcal{U}[0, A_{\max}]$, with $A_{\max}$ sampled from the real LC peak count rate distribution. The latter was obtained applying a fine-tuned peak-searching algorithm called {\sc mepsa} \citep{Guidorzi15a} to the GRB LCs that satisfied the original constraints (see Sect.~\ref{ss:lc_selection}). In this work, $A$ is sampled from the following broken power-law (BPL) probability density function (PDF):
\begin{equation}
    p_F(F) =  
    \begin{cases}
        p_0\,\left(F/F_{\mathrm{break}}\right)^{-\alpha_{\mathrm{BPL}}} ,  & \text{for $F_{\min} \leq F < F_{\mathrm{break}}$} \\
        p_0\,\left(F/F_{\mathrm{break}}\right)^{-\beta_{\mathrm{BPL}}}  ,  & \text{for}~ F \ge F_{\mathrm{break}}
    \end{cases}\quad,
\label{eq:BPL}
\end{equation}
where $F$ is the peak flux. Sampling $A$ from Eq.~(\ref{eq:BPL}) allows us to investigate, for the first time, the peak flux distribution of the individual pulses. This PDF is characterised by the following parameters:
\begin{itemize}
    \item $\alpha_{\mathrm{BPL}}$: low-flux index;
    \item $\beta_{\mathrm{BPL}}$: high-flux index;
    \item $F_{\mathrm{break}}$: break flux;
    \item $F_{\min}$: minimum flux.
\end{itemize}
The distribution in Eq.~(\ref{eq:BPL}) is inspired by the differential version of the $\log N - \log S$ of GRBs, which counts the number of events $N(>S)$ having fluence $>S$.
The normalisation constant $p_0$ is calculated as
\begin{equation}
    p_0 = \left\{\dfrac{F_{\mathrm{break}}}{1-\alpha}\left[1-\left(\dfrac{F_{\min}}{F_{\mathrm{break}}}\right)^{1-\alpha}\right] - \dfrac{F_{\mathrm{break}}}{1-\beta}\right\}^{-1} \quad,
    \label{eq:p0}
\end{equation}
which ensures $\int_{F_{\min}}^{\infty} p_F(F) \, \mathrm{d}F = 1$. In principle, for $F_{\mathrm{break}} \la F_{\min}$, the BPL reduces to a power-law (PL) distribution. The possibility of different peak flux distributions for different datasets is meant to account for the different detected GRB populations, which in turn result from different passbands, effective areas, and trigger algorithms. For example, Swift/BAT detects softer events and higher-redshift GRBs than BATSE \citep{Band06}. 

\begin{figure*}
    \centering
    \includegraphics[width=\textwidth]{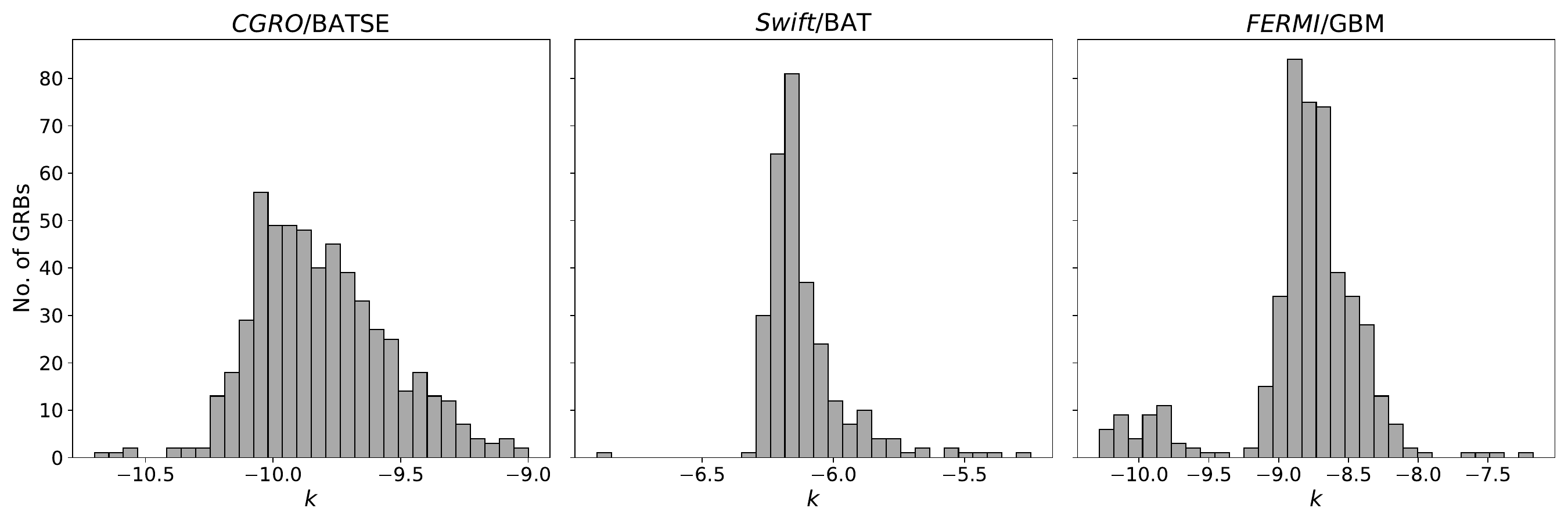}
    \caption{Distributions of the logarithmic peak-flux-to-count-rate conversion factors $k$ for \textit{CGRO}/BATSE, Swift/BAT, and Fermi/GBM.}
    \label{fig::k_factor_distr}
\end{figure*}

Peak flux values, $F$, are then converted to peak count rates or peak counts (integrated over 64~ms), $R$,  depending on the instrument via a conversion factor $k = \log_{10}(F/R)$. Here, $F$ is in units of erg~cm$^{-2}$~s$^{-1}$, while $R$ is in counts for BATSE, counts~s$^{-1}$ per fully illuminated detector for an equivalent on-axis source for Swift/BAT, and counts per detector for Fermi/GBM. For each instrument, the $k$ factors were derived as follows. We first took the fluence values (in cgs units) from the official GRB catalogues: the fourth BATSE GRB catalogue \citep{Paciesas99}, the third Swift/BAT GRB catalogue \citep{Lien16}, and the fourth Fermi/GBM GRB catalogue \citep{vonKienlin20}. Secondly, the corresponding total counts were determined by integrating over the $7\sigma$ time interval of each GRB LC. This is defined as the time interval from the first to the last bin, where the signal exceeds the background by $\ge7\sigma$ significance. The $k$ factors were then calculated as the ratio between the corresponding fluences and total counts. The distributions are shown in Fig.~\ref{fig::k_factor_distr} and reflect the variety of the GRB spectral properties.

The core of the pulse avalanche model depends on seven parameters:
\begin{itemize}
    \item $\mu_0$: the number of parent pulses in each GRB is Poisson distributed, with $\mu_0$ being the mean value;
    \item $\mu$:  the number of child pulses generated by each parent is Poisson distributed, with $\mu$ being the mean value;
    \item $\alpha$:  it rules the time delay between child and parent pulses, which is exponentially distributed;
    \item $\tau_{\min}$ and $\tau_{\max}$:  lower and upper bounds for the time constant $\tau_0$ of spontaneous pulses;
    \item $\delta_1$ and $\delta_2$: lower and upper bounds for the uniform distribution governing $\log(\tau/\tau_1)$, where $\tau$ and $\tau_1$ are the time constants of child and parent pulses, respectively.
\end{itemize}
Overall, the model consists of the combination of these seven parameters and the four BPL ones, totalling eleven parameters to be optimised by the GA (Sect.~\ref{ss:ga}).

Eventually, statistical noise is added to the simulated LCs, depending on the counts in each time bin. Following \citetalias{Bazzanini24}, background rates were set as constants for BATSE and GBM, derived from median measured error rates: 2.9~cnts~s$^{-1}$~cm$^{-2}$ for BATSE and 39.4~cnts~s$^{-1}$~cm$^{-2}$ for GBM. Final LCs are realisations of Poisson processes with expected values set equal to the sum of background and noise-free LCs, with the background subsequently removed. For BAT, the value of each bin was sampled from a Gaussian distribution with expected value equal to the noise-free LC and standard deviation taken from the distribution of the real LC error rates.

\subsection{Genetic algorithm and model parameter optimisation}
\label{ss:ga}

The code described in \citetalias{Bazzanini24}\footnote{\url{https://github.com/LBasz/geneticgrbs}} has been updated to incorporate the aforementioned changes and used to optimise the eleven model parameters. This code relies on a class of machine learning algorithms known as GAs, which mimic the natural process of biological evolution.

GAs operate by improving the average fitness of a population of solutions to an optimisation problem based on specific metrics. Over successive generations, the population evolves, eventually converging towards the optimal solution. Each solution, or `individual', is encoded as a string of parameters analogous to chromosomes in biology. The fitness of these individuals determines which ones will `mate', combining their genetic information to produce the offspring. Processes analogous to biological evolution, such as crossover (random exchange of genetic material between parents) and mutation (random alteration of a gene's value), are implemented to maintain diversity in the population. The algorithm iterates for a preselected number of generations (see \citetalias{Bazzanini24} for a detailed discussion).

In this study, each individual is represented by a sequence of values for the eleven model parameters (Sec.~\ref{ss:model}), referred to here as 'genes'. The allowed ranges of the parameters are detailed in Table~\ref{table::parameterrange}.
\begin{table}
    \caption{Region of exploration during the GA optimisation of the eleven model parameters.}
    \label{table::parameterrange} 
    \centering 
    \begin{tabular}{c c c}
        \hline\hline
        Parameter                                      & Lower bound & Upper bound \\ 
        \hline
        $\mu$                                          & $0.80$      & $1.7$ \\ 
        $\mu_0$                                        & $0.80$      & $1.7$ \\
        $\alpha$                                       & $1$         & $15$ \\
        $\delta_1$                                     & $-1.5$      & $-0.30$ \\
        $\delta_2$                                     & $0$         & $0.30$ \\
        $\tau_{\min}~(\si{s})$                         & $0.01$      & $\texttt{bin\_time}$ \\
        $\tau_{\max}~(\si{s})$                         & $1$         & $65$ \\ 
        $\alpha_{\mathrm{BPL}}$                        & $1$         & $2$ \\
        $\beta_{\mathrm{BPL}}$                         & $2$         & $3$ \\
        $F_{\mathrm{break}}~(\si{erg~cm^{-2}~s^{-1}})$ & $10^{-7}$   & $10^{-5}$ \\
        $F_{\min}~(\si{erg~cm^{-2}~s^{-1}})$           & $10^{-8}$   & $10^{-7}$ \\
        \hline 
        \hline 
    \end{tabular}
\end{table}

The sequence of steps followed by the algorithm is the same as in \citetalias{Bazzanini24}. We here summarise the key features:
\begin{itemize}
    \item The algorithm runs for 30 generations, with a population of $N_{\mathrm{pop}} = 2000$ individuals, each represented as a string of the eleven model parameters.
    \item For each individual, $N_{\mathrm{grb}} = 2000$ GRB LCs are simulated based on the parameter values. These LCs are subjected to the same selection criteria applied to the real ones (Sec.~\ref{ss:lc_selection}). Curves that fail to meet the criteria are discarded.
    \item The algorithm evaluates the five metrics (Sec.~\ref{ss:metrics}) on the selected LCs for each individual. For the first four metrics, these results are compared with the corresponding metrics from real datasets by calculating the $L_2$ loss. For the fifth metric, a two-population Kolmogorov-Smirnov test is performed. The resulting $p$-value, $p$, is used to calculate a linearly decreasing (in $\log_{10} p$) piecewise loss function, which returns zero whether $p > \alpha = 0.05$. The final loss for each individual is defined as the average of the five metric losses, while the fitness score is the inverse of this value.
    \item At each generation, individuals are ranked based on their loss, and new offspring are generated by combining the genes of the fittest individuals (top 15\%). The algorithm also incorporates random genetic mutations, that is a gene's value is randomly sampled from its defined range rather than being inherited from the parents, with each gene having a 4\% probability of mutation.
\end{itemize}

\section{Results}
\label{sec:results}

\subsection{Optimal parameter values}

The optimised parameter values for the BATSE, Swift, and Fermi datasets are presented in Table~\ref{table::parameterresults}. To be consistent with \citetalias{Bazzanini24}, these values are determined as the median of the entire population in the final generation. The loss for the best parameter set is then evaluated on a newly generated sample of 5000 LCs. A comparison between the parameters obtained in \citetalias{Bazzanini24} and the new ones reveals two key results: (i) a further reduction in total loss is achieved in comparison with both the \citetalias{SternSvensson96} guess and \citetalias{Bazzanini24} results, and (ii) the original insight by \citetalias{SternSvensson96} remains valid, as the optimised values of $\mu$ remain close to unity. Additionally, the index $\beta_{\si{BPL}}$ is close to $5/2$ across the three datasets, which corresponds to the value of $3/2$ observed in the hard tail of the cumulative $\log N-\log S$ of GRBs and expected for a spatially homogeneous population in a locally Euclidean universe.

\begin{figure*}
   \centering
   \includegraphics[width=1\linewidth]{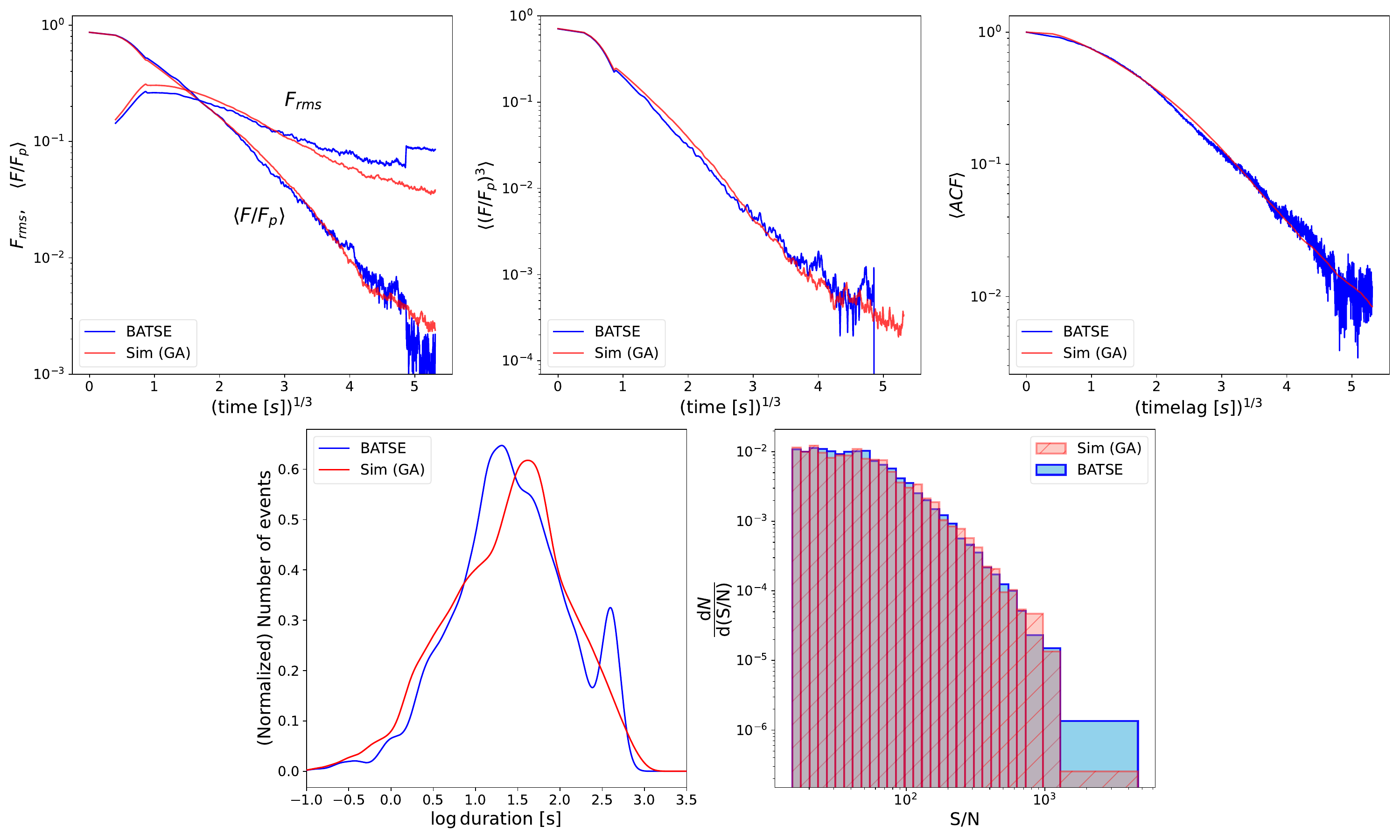}
   \caption{Distributions of the five metrics for real BATSE light curves (blue) and simulated profiles (red) obtained using the latest optimised parameter set (see Table~\ref{table::parameterresults}). \textit{Top left}: average peak-aligned post-peak normalised time profile and root mean square (r.m.s.) deviation of the individual peak-aligned profiles, $F_{\mathrm{rms}} \equiv\big[\langle(F / F_p)^2\rangle-\langle F / F_p\rangle^2\big]^{1/2}$. \textit{Top middle}: average peak-aligned third moment profiles. \textit{Top left}: average autocorrelation function of the GRBs. \textit{Bottom left}: distribution of the $T_{20\%}$ duration. \textit{Bottom right}: S/N distribution of the GRBs.
   Like in \citetalias{Bazzanini24}, the curves in the top left and top middle panels were smoothed using a Savitzky-Golay filter to reduce the effects of Poisson noise, while the distributions in the bottom left panel were smoothed with a Gaussian kernel convolution.
   }
   \label{fig::5observables_batse}
\end{figure*}

Figures~\ref{fig::5observables_batse}, \ref{fig::5observables_swift}, and \ref{fig::5observables_fermi} show the comparisons of the distributions for the five metrics between real and simulated datasets for BATSE, \emph{Swift}, and \emph{Fermi}, respectively. The compatibility between the real and simulated distributions is evident across all three datasets. Notably, the introduction of the S/N distribution metric has significantly improved the alignment between the durations of real and simulated LCs compared to previous results.

\subsection{Peak flux distributions of simulated individual pulses}
\label{ss:peak_flux_distr}

To further explore how the model parameters depend on the different effective areas, energy passbands, and trigger algorithms of BATSE, Swift/BAT, and Fermi/GBM, we compared their simulated optimal peak flux distributions $p_F(F)$ described by Eq.~(\ref{eq:BPL}). Since these were originally computed in the energy passbands of their respective instruments, we rescaled all of them to the BATSE passband (25--2000~keV) for a proper comparison. Here, we outline the procedure used for the Swift distribution; the same method was applied to Fermi. We randomly sampled $10^4$ peak flux values, $F_i$, from Eq.~(\ref{eq:BPL}) and multiplied each $F_i$ by a rescaling factor $R_i = S_i^{\mathrm{(BATSE)}}/S_i^{\rm (Swift)}$, ($i=1, \ldots, 10^4$). $S_i^{\mathrm{(BATSE)}}$ and $S_i^{\rm (Swift)}$ are, respectively, the fluence in the 25--2000 and 15--150~keV, calculated using the same set of spectral parameters randomly sampled from the BEST GRB set provided by \citet{Goldstein13}.

\begin{figure}
    \resizebox{\hsize}{!}{\includegraphics{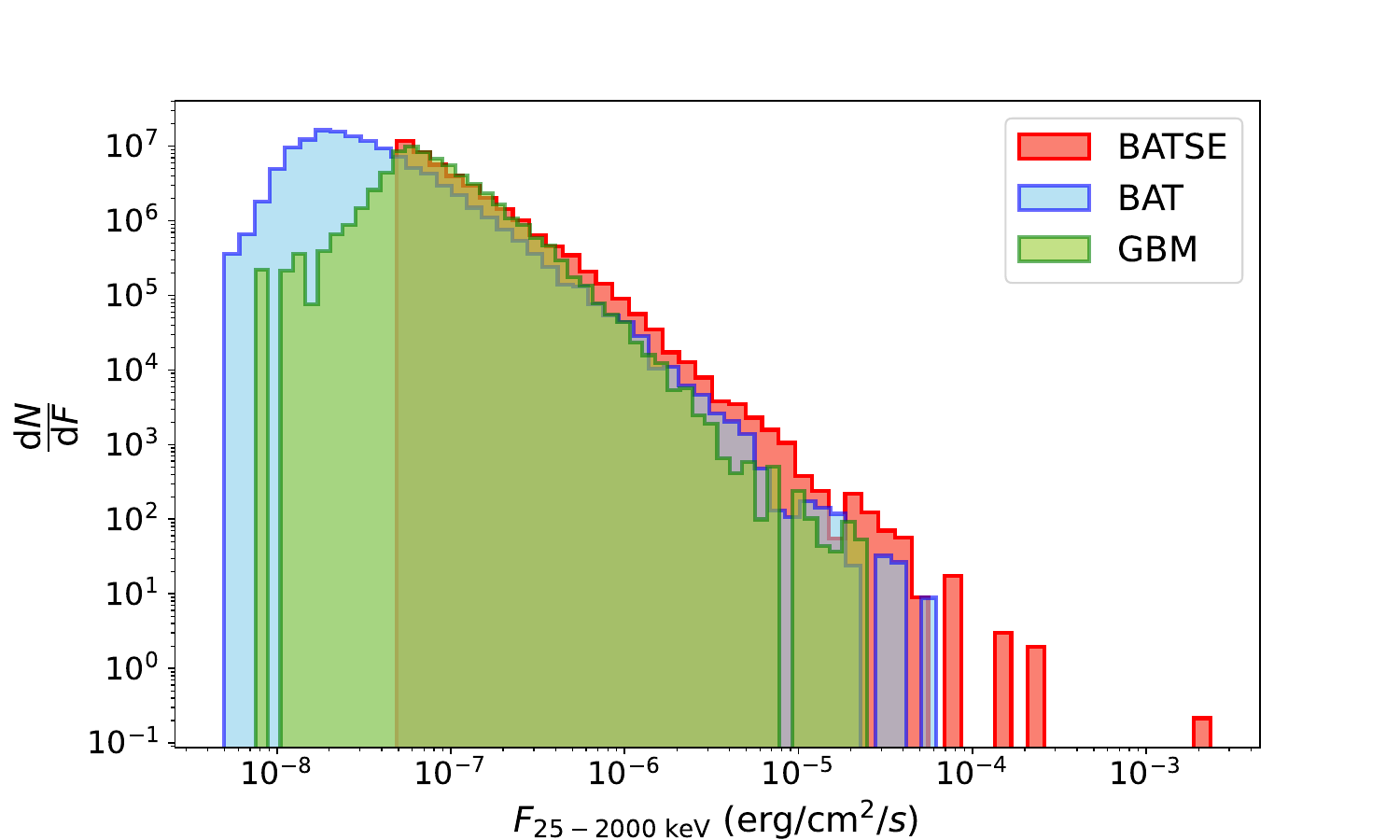}}
    \caption{Comparison between the simulated peak flux distributions of the BATSE, Swift, and Fermi samples. Values were rescaled in the BATSE energy passband (25--2000~keV). The sharp drop in the low-tail distribution of BATSE simply reflects the optimal value of the minimum peak flux $F_{min}$ obtained with the GA, whereas the same drop for the other two sets was smeared out by the rescaling in the BATSE passband.}
    \label{fig::peak_flux_distrs}
\end{figure}

As shown in Figure~\ref{fig::peak_flux_distrs}, the rescaled distributions of both Swift and Fermi overlap with the BATSE one. This indicates that the population of BATSE GRB pulses would be detectable by both BAT and GBM. Additionally, Figure~\ref{fig::peak_flux_distrs} confirms that the BAT and GBM distributions extend below that of BATSE. This result does not clash with the overall larger effective area of the latter: in fact, for soft GRBs ($E_{\rm p}\lesssim 50$--$100$~keV) the BATSE effective area significantly drops (see Figure~2 from \citealt{Tsvetkova22}), thus making  these GRBs preferably detectable by the other two instruments. These findings suggest that the differences observed in the model parameters across the three datasets can be at least partially ascribed to the GRB populations detected by each instrument.

\begingroup
\setlength{\tabcolsep}{6pt}
\renewcommand{\arraystretch}{1.4}
\begin{table*}
\caption{Optimised model parameter for the BATSE, Swift, and Fermi datasets.} 
\label{table::parameterresults}
\centering 
    \begin{tabular}{c | c | c c | c c c}
    \hline\hline
    Parameter & \citetalias{SternSvensson96}--BATSE & \citetalias{Bazzanini24}--BATSE & \citetalias{Bazzanini24}--\emph{Swift} & BATSE & Swift & Fermi\\ 
    \hline
    $\mu$                                               & $1.20$         & $1.10^{+0.03}_{-0.02}$  & $1.34^{+0.03}_{-0.02}$  & $ 0.84^{+0.03}_{-0.03}$ & $ 1.06^{+0.02}_{-0.01}$ & $ 0.97^{+0.01}_{-0.07}$\\ 
    $\mu_0$                                             & $1.00$         & $0.91^{+0.06}_{-0.07}$  & $1.16^{+0.18}_{-0.10}$  & $ 0.98^{+0.22}_{-0.1}$  & $ 1.24^{+0.20}_{-0.28}$ & $ 1.55^{+0.09}_{-0.19}$\\ 
    $\alpha$                                            & $4.00$         & $2.57^{+0.07}_{-0.52}$  & $2.53^{+0.25}_{-0.01}$  & $ 9.62^{+0.03}_{-1.24}$ & $ 7.03^{+0.00}_{-0.64}$ & $ 3.85^{+0.56}_{-0.00}$\\ 
    $\delta_1$                                          & $-0.50$        & $-1.28^{+0.16}_{-0.05}$ & $-0.75^{+0.11}_{-0.29}$ & $-1.36^{+0.16}_{-0.10}$ & $-1.37^{+0.21}_{-0.08}$ & $-0.99^{+0.20}_{-0.19}$\\ 
    $\delta_2$                                          & $0$            & $0.28^{+0.01}_{-0.03}$  & $0.27^{+0.01}_{-0.02}$  & $ 0.08^{+0.10}_{-0.04}$ & $ 0.04^{+0.04}_{-0.03}$ & $ 0.03^{+0.01}_{-0.02}$\\ 
    $\tau_{\min}~(\si{s})$                              & $0.02$         & $0.02^{+0.02}_{-0.01}$  & $0.03^{+0.02}_{-0.02}$  & $ 0.02^{+0.02}_{-0.01}$ & $ 0.02^{+0.03}_{-0.01}$ & $ 0.03^{+0.01}_{-0.01}$\\ 
    $\tau_{\max}~(\si{s})$                              & $26.0$         & $40.2^{+0.9}_{-1.2}$    & $56.8^{+0.4}_{-1.3}$    & $33.15^{+0.06}_{-1.29}$ & $62.46^{+0.19}_{-2.82}$ & $35.84^{+0.87}_{-0.00}$\\
    $\alpha_{\mathrm{BPL}}$                             & --             & -- & -- & $ 1.61^{+0.00}_{-0.11}$ & $ 1.89^{+0.03}_{-0.01}$ & $ 1.88^{+0.03}_{-0.09}$\\
    $\beta_{\mathrm{BPL}}$                              & --             & -- & -- & $ 2.19^{+0.08}_{-0.05}$ & $ 2.53^{+0.11}_{-0.10}$ & $ 2.58^{+0.06}_{-0.11}$\\
    $F_{\si{break}}~(10^{-7}~\si{erg~cm^{-2}~s^{-1}})$  & --             & -- & -- & $ 6.18^{+2.20}_{-0.20}$ & $ 3.44^{+0.10}_{-0.60}$ & $ 2.88^{+0.70}_{-0.20}$\\
    $F_{\min}~(10^{-8}~\si{erg~cm^{-2}~s^{-1}})$        & --             & -- & -- & $ 4.87^{+1.16}_{-0.52}$ & $ 1.41^{+0.06}_{-0.04}$ & $ 6.04^{+0.48}_{-0.44}$\\
    \hline 
    Loss (\textit{Train} best)                          & --      & $0.72$ & $0.38$ & $0.38$ & $0.46$ & $0.54$\\
    Loss (\textit{Train} avg.)                          & --      & $0.98$ & $0.66$ & $0.84$ & $0.89$ & $0.94$\\
    \hline 
    Loss (\textit{Test})                                & $1.47$  & $0.88$ & $0.56$ & $0.49$ & $0.55$ & $0.61$\\
    \hline 
    Loss (\textit{Test}: $\langle F / F_p\rangle$)      & $1.01$  & $0.67$ & $0.46$ & $0.19$ & $0.40$ & $0.13$\\
    Loss (\textit{Test}: $\langle(F / F_p)^3\rangle$)   & $0.40$  & $0.20$ & $0.20$ & $0.27$ & $0.28$ & $0.18$\\
    Loss (\textit{Test}: $\langle\mathrm{ACF}\rangle$)  & $2.24$  & $0.64$ & $0.49$ & $0.38$ & $0.71$ & $0.58$\\
    Loss (\textit{Test}: $T_{20\%}$)                    & $2.22$  & $2.04$ & $1.08$ & $1.62$ & $1.37$ & $2.17$\\
    Loss (\textit{Test}: S2N distr.)                    & --      & --     & --     & 0      &      0 &      0\\
    \hline 
    \hline 
    \end{tabular}
    \tablefoot{Column~2 lists the parameters provided by \citetalias{SternSvensson96} for the BATSE dataset, while Columns~3 and~4 present the optimised parameters provided by \citetalias{Bazzanini24} for BATSE and Swift/BAT, respectively. Columns~5, ~6, and~7 report the optimised parameters for BATSE, Swift/BAT, and Fermi/GBM, respectively, obtained after 30 generations of the GA. The best-fitting values of the 11 parameters were determined as the medians of their distributions in the final generation, with uncertainties estimated from the 16th and 84th percentiles. "Train best" refers to the loss of the best-performing generation, while "Train avg." represents the average loss in the final generation. The test set consists of a newly generated collection of 5000 simulated LCs, with the last five rows detailing the individual contributions to the "Test" loss.
    }
\end{table*}
\endgroup

\section{Discussion and conclusions}
\label{sec:disc_conc} 

\citetalias{Bazzanini24} first used a GA to optimise the parameters of the stochastic pulse avalanche model originally proposed by \citetalias{SternSvensson96}, by simulating LCs that replicate some of the observed average properties of real LCs of LGRBs detected by BATSE and Swift/BAT. In this work, we built on the \citetalias{Bazzanini24} results by introducing an additional metric --the S/N distribution of GRBs-- and incorporating four additional model parameters that characterise the flux distribution of individual pulses in terms of a BPL.

As in \citetalias{Bazzanini24}, we tested our advanced model against the BATSE and Swift/BAT catalogues. Our results significantly improved, as attested by the reduced loss of the four previously defined metrics along with a low loss on the fifth one (Table~\ref{table::parameterresults}). Additionally, the model was tested for the first time on a complementary dataset, the Fermi/GBM catalogue, yielding analogously satisfactory results. Notably, for all three datasets the parameter $\mu$, which governs the number of child pulses generated by each parent, systematically converged to unity. This outcome further supports the hypothesis that LGRB central engines, or, more generally, the source of variability driving the dissipation mechanism, operates close to a critical regime. As mentioned in \citetalias{Bazzanini24}, if variability originates from the central engine, as suggested by the internal shock model \citep{Rees94, Kobayashi97, Daigne98, Maxham09}, the branching nature of the avalanche process could stem from fragmentation driven by magneto-rotational, gravitational, and/or viscous instabilities in the accretion disc of a hyper-accreting BH. Alternatively, the avalanche mechanism could result from magnetic energy dissipation through runaway sequences of reconnection events, as predicted in the ICMART model \citep{ICMART}.

Moreover, for the first time we proved that the peak flux distribution of individual GRB pulses, which, to our knowledge, has never been studied so far, can successfully be modelled as a BPL and constrained under the assumption of a relatively simple stochastic model like the one discussed in this work. Interestingly, the BPL shape of the peak flux distribution, which works for all three datasets, is reminiscent of the BPL distribution of the isotropic-equivalent peak luminosity of individual pulses, $L_{\mathrm{iso}}$, found by \citet{Maccary24}, although the latter is an intrinsic (that is, redshift-independent) property of GRBs, unlike the peak flux.

We also compared the flux distributions of individual pulses for the three datasets, rescaled to the BATSE energy passband, highlighting the influence of different instrumental properties, such as passband, effective area, and trigger algorithm, on the model parameters. These results propel the algorithm developed in this work as a robust and reliable tool for simulating realistic LGRB LCs for upcoming experiments, for example HERMES \citep{Fiore20_HERMES}, and future missions, such as the X/Gamma-ray Imaging Spectrometer (XGIS; \citealp{Amati22}) aboard the EAS/M7 candidate THESEUS \citep{Amati21b}, using parameters tailored to the instrumental characteristics of each experiment.

The source code for our algorithm, along with all scripts used for data analysis and visualisation, is publicly available on GitHub\footnote{\url{https://github.com/mmanuele99/geneticgrbs_v2}}.

\begin{acknowledgements}
We acknowledge the reviewer for their useful insights and comments, which helped us to improve the clarity of the manuscript. This work uses the following software packages:
    \href{https://www.python.org/}{\texttt{Python}}
    \citep{python},
    \href{https://github.com/ahmedfgad/GeneticAlgorithmPython}{\texttt{PyGAD}}
    \citep{gad2023pygad},
    \href{https://github.com/numpy/numpy}{\texttt{NumPy}}
    \citep{numpy1, numpy2},
    \href{https://github.com/scipy/scipy}{\texttt{SciPy}}
    \citep{scipy},
    \href{https://github.com/matplotlib/matplotlib}{\texttt{matplotlib}}
    \citep{matplotlib},
    \href{https://www.fe.infn.it/u/guidorzi/new_guidorzi_files/code.html}{\sc{mepsa}}
    \citep{Guidorzi15a},
    \href{https://www.gnu.org/software/bash/}{\texttt{bash}}
    \citep{gnu2007free}.
    
\end{acknowledgements}

\bibliographystyle{aa}
\bibliography{alles_grbs}

\begin{appendix}
\onecolumn
\section{Comparisons of the five metrics between real and simulated datasets}

\begin{figure*}[h!]
   \centering
   \includegraphics[width=0.9\linewidth]{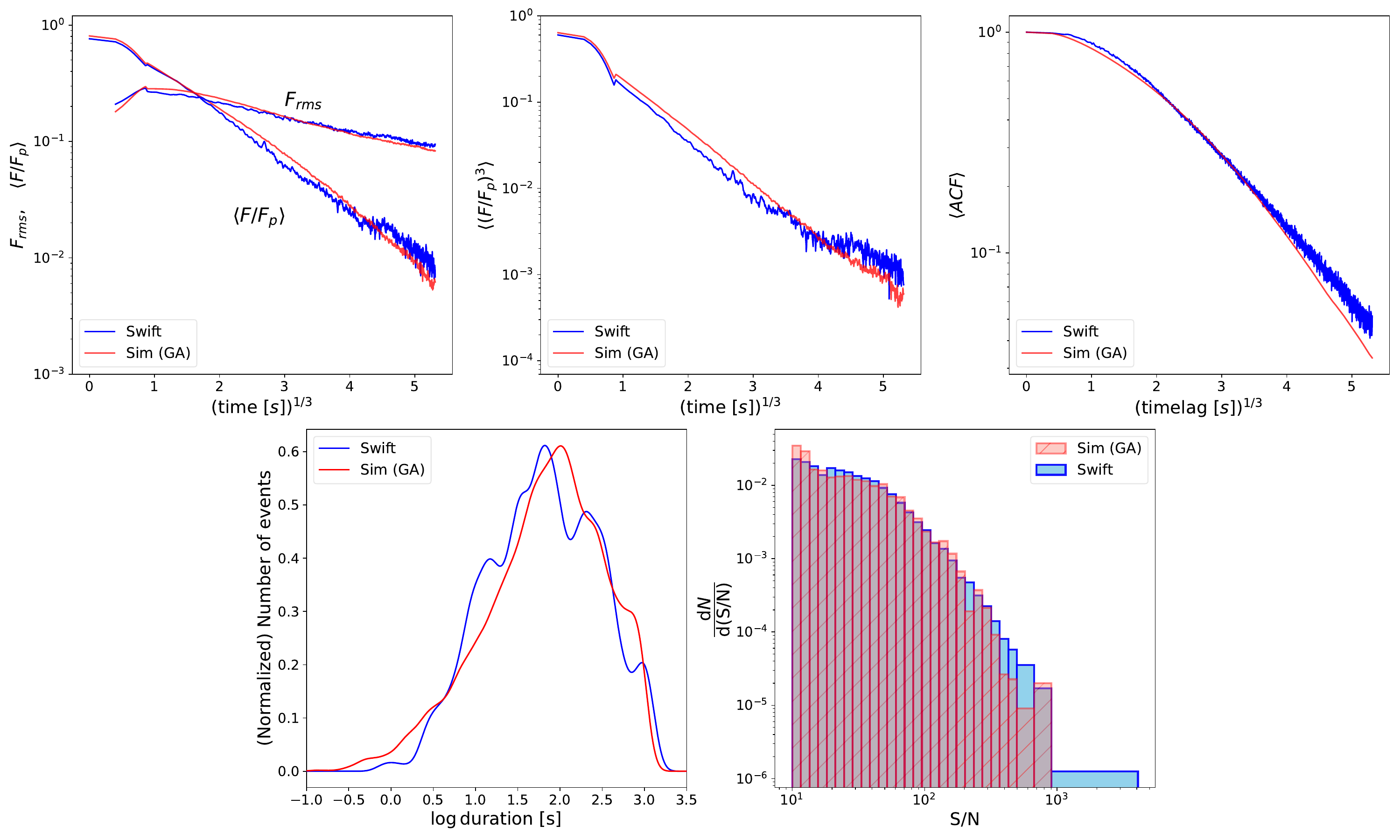}
   \caption{Comparison between the real Swift/BAT dataset and the corresponding simulated dataset on the same metrics defined for the BATSE dataset, as in Fig.~\ref{fig::5observables_batse}.}
   \label{fig::5observables_swift}
\end{figure*}

\begin{figure*}[h!]
   \centering
   \includegraphics[width=0.9\linewidth]{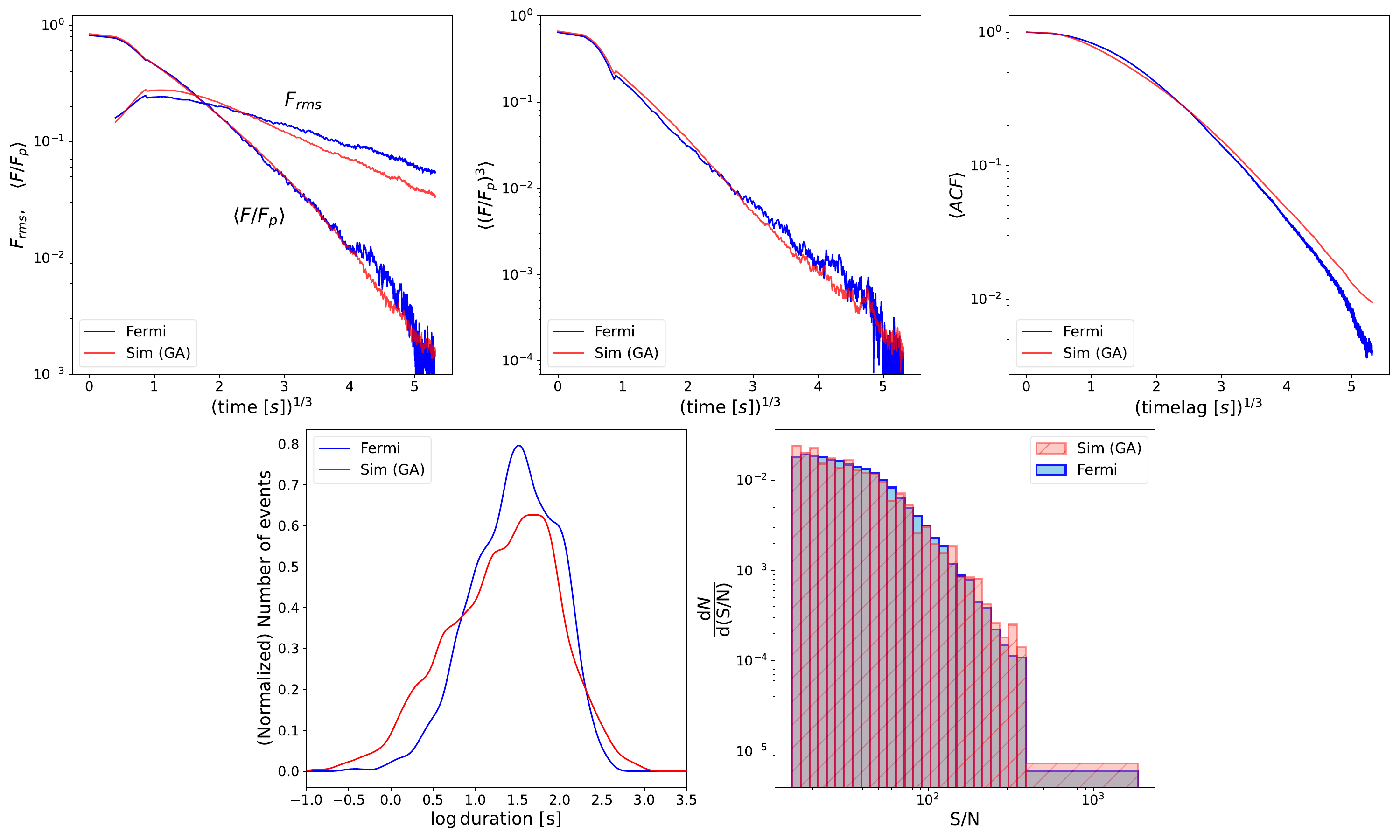}
   \caption{Comparison between the real Fermi/GBM dataset and the corresponding simulated dataset on the same metrics defined for the BATSE dataset, as in Fig.~\ref{fig::5observables_batse}.}
   \label{fig::5observables_fermi}
\end{figure*}
\end{appendix}

\end{document}